\documentclass[preprint,showpacs,preprintnumbers,amsmath,amssymb]{revtex4}
\usepackage{graphicx}
\usepackage{amssymb}

\def\undersim#1{\setbox9\hbox{${#1}$}{#1}\kern-\wd9\lower
    2.5pt \hbox{\lower\dp9\hbox to \wd9{\hss $_\sim$\hss}}}
\def\bqr{{\hbox{\boldmath $\rho$}}}

\begin{document}
\title{Fluctuations of pion elliptic flow, triangular flow, and HBT correlation functions in
ultrarelativistic heavy ion collisions}

\author{Ying Hu$^1$, Wei-Ning Zhang$^{1,2}$\footnote{Email: wnzhang@dlut.edu.cn}, Yan-Yu Ren$^2$}
\affiliation{
$^1$School of Physics and Optoelectronic Technology, Dalian University of Technology,
Dalian, Liaoning $116024$, China\\
$^2$Department of Physics, Harbin Institute of Technology, Harbin, Heilongjiang 150006, China
}

\begin{abstract}
We investigate the fluctuations of pion elliptic flow, triangular flow, and Hanbury-Brown-Twiss
(HBT) correlation functions for the hydrodynamic sources with fluctuating initial conditions in
the heavy ion collisions of the Au-Au at $\sqrt{s_{NN}}=200$ GeV and the Pb-Pb at $\sqrt{s_{NN}}
=2.76$ TeV.  A method based on event-subcollection analysis is used to detect these fluctuations
in ultrarelativistic heavy ion collisions.  We introduce a granularity length to describe the
granular inhomogeneity of the initial sources, and investigate its relationships with the
fluctuations of the flow harmonics and HBT correlation functions.  Our investigations indicate
that the fluctuations of the triangular flow of event subcollections are sensitive to the
granularity length of the initial source.  This dependence provide a way to investigate the 
granular inhomogeneity of the initial source through analysing the fluctuations of triangular 
flow in ultrarelativistic heavy ion collisions.  
\end{abstract}
\pacs{25.75.-q, 25.75.Gz, 25.75.Ld}
\maketitle

\section{Introduction}
The primary goal of the heavy ion collisions at the highest energy of the Relativistic
Heavy Ion Collider (RHIC) and the energy of the Large Hadron Collider (LHC) is to explore
the properties of the extreme hot and dense matter, the quark-gluon plasma (QGP), existed
in the early stages of the collisions.  Recent investigations \cite{ISF1214,Sch12} indicate 
that the initial systems created in the ultrarelativistic heavy ion collisions at the RHIC 
and LHC energies are not uniform in space, and there are event-by-event fluctuations of the 
system initial quantities, due to the fluctuations of nucleon distributions in the nuclei, 
the fluctuations in the color charge distributions inside a nucleon, and combined with 
highly Lorentz contraction.  The studies of the system evolution with the fluctuating
initial conditions (FIC) and the influence of the FIC on final particle observables 
are recently very interesting issues in high energy heavy ion collisions \cite{ISF1214}.  
They are important to improve our understanding of the experimental results at the RICH 
and the LHC.

Elliptic flow and two-particle Hanbury-Brown-Twiss (HBT) correlation functions are important
observables in high energy heavy ion collisions.  They reflect the transverse (perpendicular
to the beam direction) anisotropic pressure property and the space-time structure of the
particle-emitting sources, respectively.  In ultrarelativistic heavy ion collisions, the
spectators depart from the reaction region quickly after collision, and a very hot and 
dense fireball is created in the mid-rapidity region.  For uniform systems of the fireballs 
the odd-order azimuthal flow harmonics are expected to be zero.  However, recent studies 
indicate that the fluctuating inhomogeneous density distributions of the initial systems 
may lead to nonzero triangular flow, and thus inspires the investigations of azimuthal 
triangular flow and even higher-order flow harmonics \cite{{ISF1214,Sch12,AlvRol10,Alv10a, 
StaShu11,Sch11,vn-exp,Gal13,LXHan11,LMa14,Bra14}}.  In Ref. \cite{Gal13}, the authors 
reproduced well the experimental flow results of $v_2$, $v_3$, $v_3$, and $v_5$ in the 
ultrarelativistic heavy ion collisions at the RHIC and the LHC, by using the IP-Glasma 
FIC \cite{Sch12} with the viscous hydrodynamic model of MUSIC \cite{Sch11}.  The 
investigations imply that the fluctuations in the initial geometry state are important 
and the created medium behaves as a nearly perfect liquid of nuclear matter because it 
has an extraordinarily low ratio of shear viscosity to entropy density.  

On event-by-event basis, the particle-emitting sources with the FIC are bumpy and 
inhomogeneous \cite{Gyu97,Osa02,RenZha08,Wer10,Sch11,HuZha13}.  This inhomogeneous 
structure may lead to the fluctuations of final observables event-by-event.  Although 
the influence of the IFC can be analysed by some observables, for instance, nonzero 
$v_3$, our motivation here is to detect the fluctuations of final observables directly 
and try to look for the relationship between the fluctuations of final observables and 
the granular inhomogeneity of the initial sources.

In this paper, we use the Heavy Ion Jet Interaction Generator (HIJING) \cite{Wan05} to 
generate the FIC of the particle-emitting sources for the heavy ion collisions of the 
Au-Au at $\sqrt{s_{NN}}=200$ GeV at the RHIC and the Pb-Pb at $\sqrt{s_{NN}}=2.76$ TeV 
at the LHC.  The system evolution is described by the relativistic ideal hydrodynamics 
in (2+1) dimensions with the Bjorken longitudinal boost invariance \cite{Bjo83}, and 
with the equation of state (EOS) of s95p-PCE \cite{She10}.  We use the relativistic 
Harten-Lax-Leer-Einfeldt (RHLLE) algorithm \cite{Ris98,HLLE,Ris9596,ZhaEfa,Zha04,Yu08Yin12} 
in our hydrodynamic calculations.  Although this algorithm is also valid for the viscous 
hydrodynamic source with rest and smoothed initial conditions \cite{Efa12CPC}, it is hard 
to obtain the stable numerical solutions for the viscous hydrodynamics with the FIC and 
nonzero initial velocities of fluid-cells in the source with the RHLLE algorithm.  
We calculate the pion elliptic flow, triangular flow, and HBT correlation functions for 
the ideal hydrodynamic sources with the FIC and the nonzero initial fluid velocity.  
Motivated by the works on the fluctuations of single-event HBT correlation functions for
granular sources \cite{WonZha04,Zha05,RenZha08,HuZha13}, we investigate the fluctuations 
of the flow harmonics and HBT correlation functions of event subcollections for the FIC 
sources.  We introduce a granularity length to describe the granular inhomogeneity of
the initial sources and investigate its relationships with the fluctuations of the flow 
harmonics and HBT correlation functions.  Our investigations indicate that the FIC lead 
to event-by-event fluctuations of the elliptic flow, triangular flow, and HBT correlation 
functions.  These FIC-caused fluctuations can be detected by the fluctuation distributions 
of the observables of event subcollections.  The fluctuations of the triangular flow of 
event subcollections are sensitive to the granularity length of the initial source.  
This dependence provide a way to investigate the granular inhomogeneity of the initial 
sources through analysing the fluctuations of triangular flow in ultrarelativistic 
heavy ion collisions.  

The rest of the paper is organized as follows.  In Sec. II, we will give a description 
on solving the hydrodynamic equations in (2+1) dimensions and present the space-time 
evolutions of the sources with the FIC.  In Sec. III, we will investigate the fluctuations 
of the pion elliptic flow, triangular flow, and HBT correlation functions in the event 
subcollections for the heavy ion collisions of the Au-Au collisions at $\sqrt{s_{NN}}=200$ 
GeV at the RHIC and the Pb-Pb at $\sqrt{s_{NN}}=2.76$ TeV at the LHC.  The relationship 
between the fluctuations of the observables of event subcollections and the granular 
inhomogeneity of initial source is investigated also in Sec. III.  Finally, we will give 
the summary and discussions in Sec. IV.

\section{Hydrodynamic evolution of the sources with HIJING FIC}
The description of ideal hydrodynamics for the system with zero net-baryon density is defined
by the local conservations of energy and momentum\cite{Ris98,KolHei03},
\begin{equation}\label{hy1}
\partial_{\mu}T^{\mu\nu}=0,
\end{equation}
where $T^{\mu\nu}\!=\!(\epsilon\!+\!{\cal P})u^{\mu}u^{\nu}\!-\!{\cal P} g^{\mu\nu}$ is the
density tensor of energy-momentum of ideal fluid, $\epsilon$ and ${\cal P}$ are the energy
density and pressure in the local rest frame of the fluid element which moving with velocity
$\textbf{\emph{v}}$, $u^{\mu}=\gamma(1,\textbf{\emph{v}})$ is the four-velocity, $\gamma\!=\!
(1\!-\!\textbf{\emph{v}}^{2})^{-1/2}$, and $g^{\mu\nu}\!=\mathrm{diag}(+,-,-,-)$ is Minkowski
metric tensor.  Under the assumption of Bjorken longitudinal boost invariance \cite{Bjo83},
the hydrodynamics in (3+1) dimensions reduces to (2+1) dimensions.  In this case we need only
to solve the transverse equations of motion in $z=0$ plane, and the hydrodynamic solutions
at $z\ne0~(v^z=z/t)$ can be obtained by the longitudinal boost invariance hypothesis
\cite{Bay83,Gyu97}.

From Eq. (\ref{hy1}) we have the transverse equations in $z=0$ plane,
\begin{eqnarray}\label{hyeq2}
&&\hspace*{-8mm}\partial_t {\cal E}+\partial_x [({\cal E}\!+\!{\cal P})v^x]+\partial_y
[({\cal E}\!+\!{\cal P}) v^y]=-{\cal F}({\cal E},{\cal P},t),
\nonumber\\
&&\hspace*{-8mm}\partial_t {\cal M}^x+\partial_x({\cal M}^x v^x\!+\!{\cal P})+\partial_y
({\cal M}^x v^y)=-{\cal G}({\cal M}^x,t),\\
&&\hspace*{-8mm}\partial_t {\cal M}^y+\partial_x({\cal M}^y v^x)+\partial_y({\cal M}^y
v^y\!+\!{\cal P})=-{\cal G}({\cal M}^y,t),
\nonumber
\end{eqnarray}
where ${\cal E}=T^{00}$, ${\cal M}^i=T^{0i},\,(i=x,y)$, ${\cal F}({\cal E},{\cal P},\,t)=
({\cal E}+{\cal P})/t$, and ${\cal G}({\cal M}^i,t)={\cal M}^i\!/t$.  In equation set
(\ref{hyeq2}) there are $\epsilon$, ${\cal P}$, $v^x$, and $v^y$ four variables.  So an EOS,
${\cal P}(\epsilon)$, is needed to enclose the equation set.  In the calculations, we use
the EOS of s95p-PCE, which combines the hadron resonance gas at low temperature and the
lattice QCD results at high temperature \cite{She10}.

Assuming the local equilibrium of system is reached at time $\tau_{0}$, we construct
the initial energy density of the hydrodynamic source at $z=0$, by using the AMPT code
\cite{Lin05} in which the HIJING is used for generating the initial conditions, as
\cite{Gyu97,Pan12}
\begin{eqnarray}
\label{E-ic}
\epsilon(\tau_0,x,y;z=0)=K\sum_{\alpha}\frac{p_{\perp\alpha}}{\tau_0}\frac{1}{2\pi
\sigma_0^2}\,\exp \left\{-\frac{[x-x_{\alpha}(\tau_{0})]^{2}+[y-y_{\alpha}(\tau_{0}
)]^2}{2\pi\sigma_0^2}\right\}.
\end{eqnarray}
Here $p_{\perp\alpha}$ is the transverse momentum of parton $\alpha$ in the fluid element
at $(x,y)$, $x_{\alpha}(\tau_{0})$ and $y_{\alpha}(\tau_{0})$ are the transverse coordinates
of the parton at $\tau_{0}$, $\sigma_0$ is a transverse width parameter, and $K$ is a scale
factor which can be adjusted to fit the experimental data of produced hadrons \cite{Pan12}.
The initial velocity of the fluid element is then determined by the initial energy density
and the average transverse momentum of the partons in the element.

With the EOS and the initial values of energy density and velocities, we can solve equation
set (\ref{hyeq2}) using the relativistic HLLE scheme and Sod's operation splitting method
\cite{HLLE,Ris98,Ris9596,ZhaEfa,Zha04,Yu08Yin12,Sod77}: first getting the solutions for the 
corresponding homogeneous equations in $x$ and $y$ directions; then obtaining the solutions 
of (\ref{hyeq2}) with the corrections of ${\cal F}$ and ${\cal G}$ to the solutions of the 
homogeneous equations.  In our calculations the spatial grid sizes are taken to be $\Delta 
x=\Delta y=0.1$ fm, and the time step is taken to be $\Delta t=0.99\Delta x$ 
\cite{Ris9596,ZhaEfa,Zha04,Yu08Yin12}. \\

\begin{figure*}[!htbp]
\begin{center}
\begin{minipage}{0.16\textwidth}
\hspace*{-1mm}\includegraphics[scale=0.5]{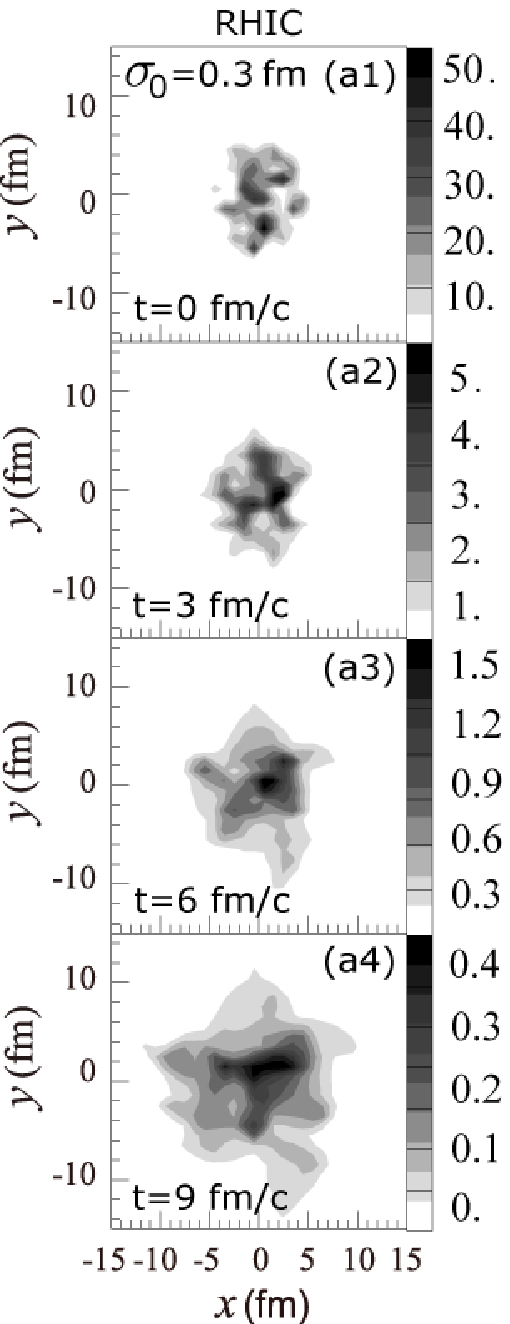}
\end{minipage}%
\begin{minipage}{0.16\textwidth}
\hspace*{-1mm}\includegraphics[scale=0.5]{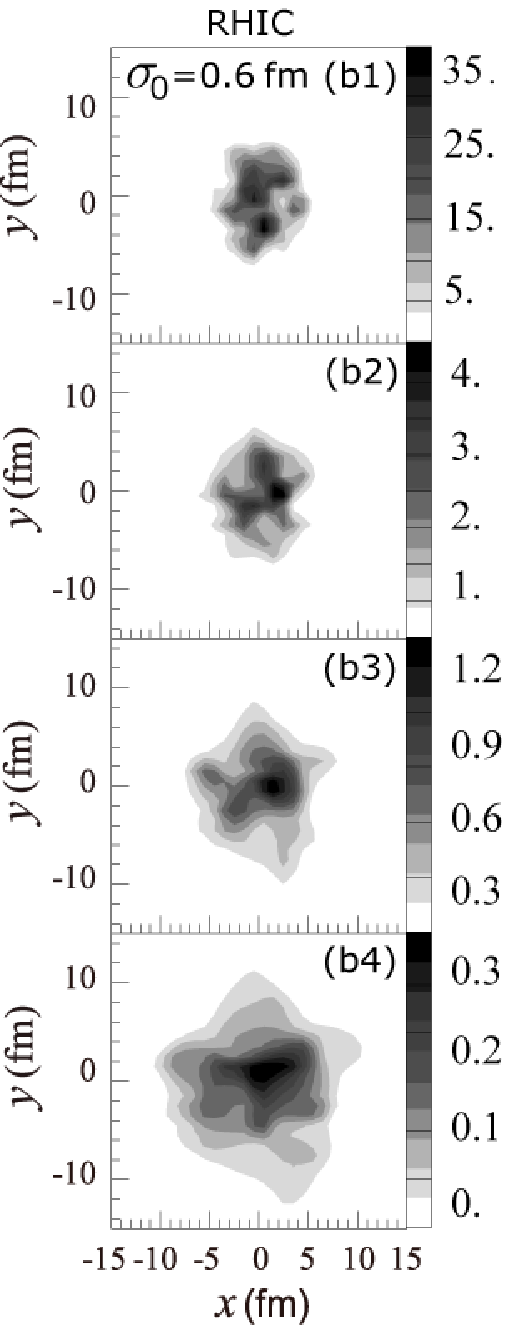}
\end{minipage}%
\begin{minipage}{0.16\textwidth}
\hspace*{0mm}\includegraphics[scale=0.5]{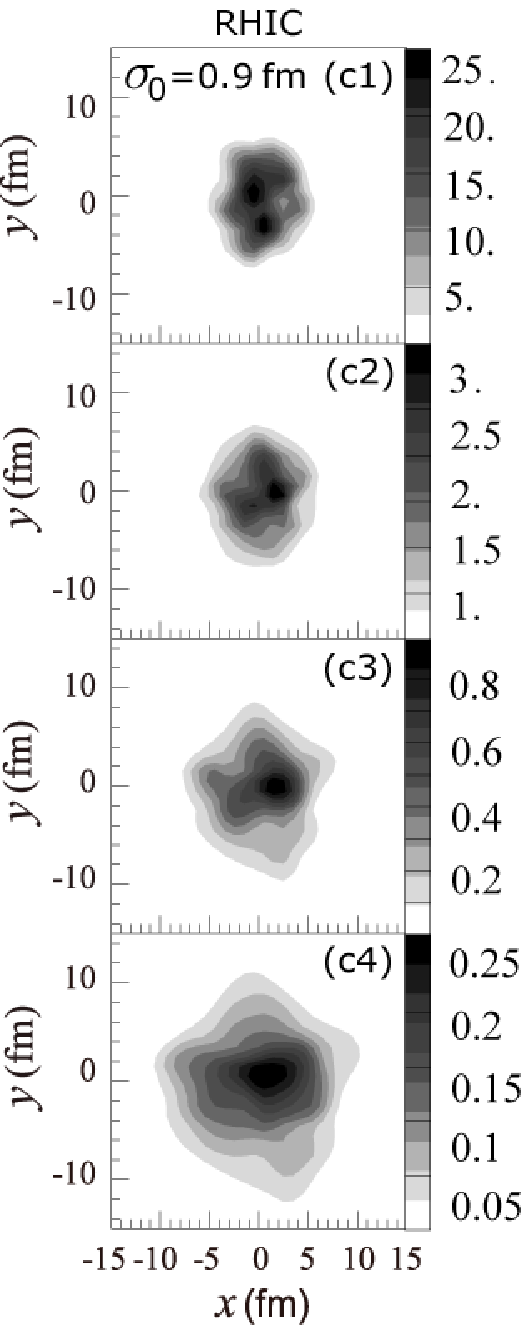}
\end{minipage}%
\begin{minipage}{0.16\textwidth}
\hspace*{1mm}\vspace*{-1mm}\includegraphics[scale=0.5]{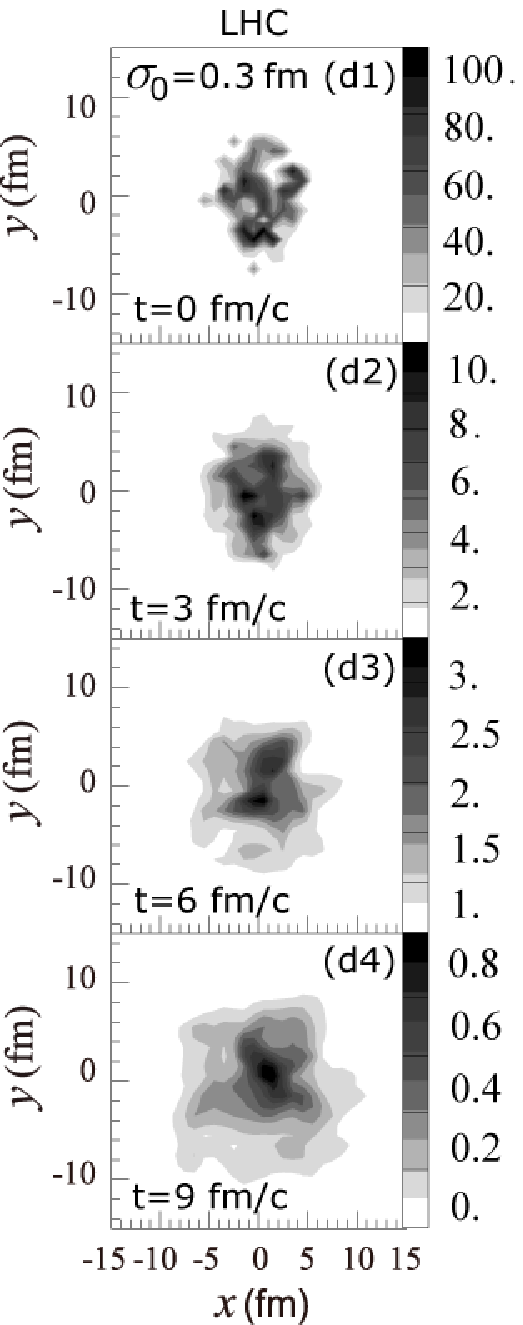}
\end{minipage}%
\begin{minipage}{0.16\textwidth}
\hspace*{1mm}\vspace*{-1mm}\includegraphics[scale=0.5]{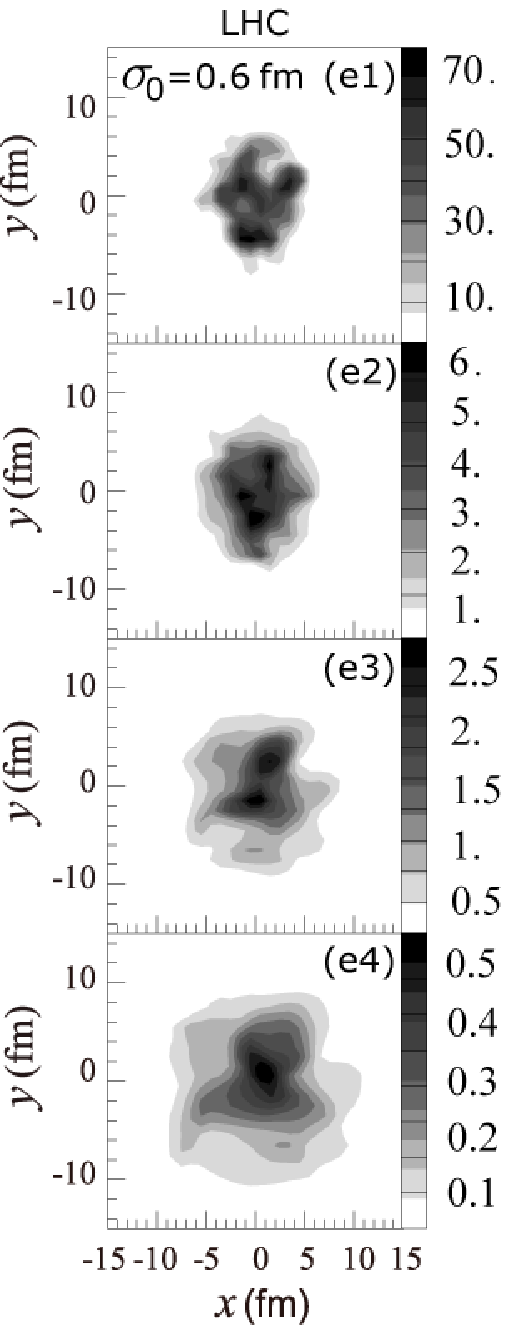}
\end{minipage}%
\begin{minipage}{0.16\textwidth}
\hspace*{1mm}\vspace*{-1mm}\includegraphics[scale=0.5]{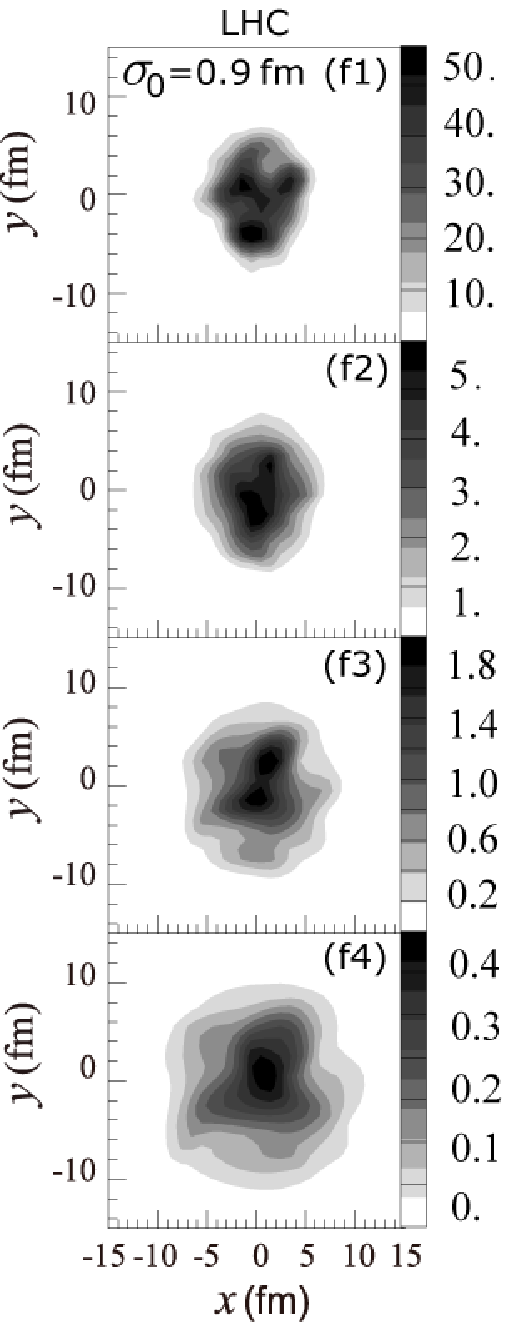}
\end{minipage}%
\caption{The transverse distributions of energy density at $z=0$ for the Au-Au collisions
at $\sqrt{s_{NN}}=200$ GeV at the RHIC and the Pb-Pb collisions at $\sqrt{s_{NN}}=2.76$
TeV at the LHC.  The impact parameter $b$ for both the RHIC and LHC heavy ion collisions
is 4 fm.  The $\sigma_0$ values are 0.3, 0.6, and 0.9 fm.  The panels [(a1)-- (f1)],
[(a2)-- (f2)], [(a3)-- (f3)], [(a4)-- (f4)] are for the evolution time $t=$ 0, 3, 6, and 
9 fm/$c$ after $\tau_0$, respectively.  The unit of energy density is GeV/fm$^3$. }
\label{zf-hydro}
\vspace*{-5mm}
\end{center}
\end{figure*}

In Figs. \ref{zf-hydro}[(a1)-- (c1)] and \ref{zf-hydro}[(d1)-- (f1)], we show the transverse
distributions of the source initial energy density at $z=0$ for the Au-Au and Pb-Pb collisions
at the RHIC and LHC energies, respectively.  Here the unit of energy density is GeV/fm$^3$.
The impact parameter $b$ for both the RHIC and LHC heavy ion collisions is 4 fm, and the
$\tau_0$ values for the RHIC and LHC collisions are taken to be 0.6 and 0.4 fm/$c$,
respectively.  The energy densities at the evolution time 3, 6, and 9 fm/$c$ after $\tau_0$
are shown in the panels [(a2)-- (a4)], [(b2)-- (b4)], [(c2)-- (c4)], [(d2)-- (d4)], [(e2)--
(e4)], and [(f2)-- (f4)] for the RHIC and LHC sources with the different $\sigma_0$,
respectively.  One can see from the panels [(a1)-- (c1)] and [(d1)-- (f1)] that the initial
energy density is fluctuated.  There are hot spots and cold valleys in the systems.
We call this inhomogeneous structure the granular inhomogeneity of the initial source.
The maximum of the energy density of spot decreases when $\sigma_0$ increases.  Also, the
spot number decreases with increasing $\sigma_0$.  From the panels of time greater than
zero one can see that the sources are still inhomogeneous at the late stages of the evolution,
due to the initial fluctuations.

\vspace*{2mm}
\begin{figure}[!htb]
\includegraphics[scale=1.2]{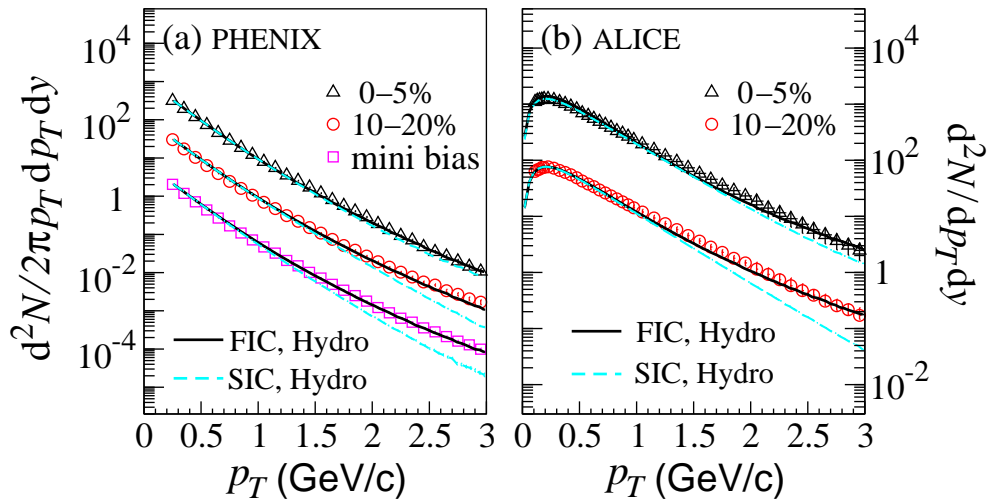}
\caption{(Color online) The pion transverse momentum spectra for the Au-Au collisions at the
RHIC energy $\sqrt{s_{NN}}=200$ GeV [panel (a)] and the Pb-Pb collisions at the LHC energy
$\sqrt{s_{NN}}=2.76$ TeV [panel(b)].  The solid lines are the hydrodynamic results with the
FIC, and the dashed lines are the hydrodynamic results with the SIC which are obtained by
averaging the FIC over 100 events.  The circle, triangle, and square symbols are the
experimental data at the RHIC \cite{PHEpt} and the LHC \cite{ALIpt}. }
\label{zf-dndpt}
\end{figure}

We show in Figs. \ref{zf-dndpt}(a) and \ref{zf-dndpt}(b) the pion transverse momentum spectra
calculated by the hydrodynamics with the FIC (solid lines) and the smoothed initial conditions
(SIC) which are obtained by averaging the FIC over 100 events, for the Au-Au and Pb-Pb collisions
at the RHIC and LHC energies, respectively.  Here, the circle, triangle, and square symbols
are the experimental data at the RHIC \cite{PHEpt} and the LHC \cite{ALIpt}.  In the hydrodynamic
calculations, we take the particle rapidity cuts the same as in the experimental analyses at the
RHIC \cite{PHEpt} and the LHC \cite{ALIpt}, respectively.  The freeze out temperature is taken
to be 130 MeV, and the parameter $\sigma_0$ is 0.6 fm.  For the centralities 0--5\%, 10--20\%,
and mini bias, the regions of impact parameter are taken to be 0--2.3, 4.2--5.9, and 0--10.2 fm,
respectively \cite{STAv2-05}.  It can be seen that the hydrodynamic results with the FIC are
consistent with the experimental data.  At large $p_T$, the spectrum of the hydrodynamic source
with the FIC is higher than the corresponding spectrum of the hydrodynamic source with the SIC.

\section{Fluctuations of pion flow harmonics and HBT correlation functions}
\subsection{Flow harmonics of event subcollections}
In high energy heavy ion collisions, the invariant momentum distribution of final particles can be
written in the form of a Fourier series \cite{SV-YZ96,AP-SV98},
\begin{eqnarray}
\label{pdis}
E\frac{d^3N}{d^3p}\!=\!\frac{1}{2\pi}\frac{d^2N}{p_Tdp_Tdy}\!\bigg[1+\sum_n{2v_n\cos(n\phi-n\Psi_R)}
\bigg],
\end{eqnarray}
where $\phi$ is the azimuthal angle of the particle and $\Psi_R$ is the azimuthal angle of event
reaction plane.  The first term on the right side of Eq. (\ref{pdis}) is the transverse momentum
spectrum, and the coefficients in the summation, $v_n=\langle \cos[n(\phi-\Psi_R)]\rangle$, are
the azimuthal $n$th-order flow harmonics, where $\langle \dots \rangle$ denotes the average over
the particles and events.  In experimental data analyses, the reaction plane is usually replaced
by the event plane which is determined with the measured particles in an event
\cite{PHEv2-03,Vol08}.  An alternative technique in flow analyses is the measurement of the
two-particle cumulant of azimuthal correlations, $[v_n\{2\}]^2=\langle \cos[n(\phi_1-\phi_2)]
\rangle$ \cite{Bor01,ALIv2-10}, which avoids the uncertainty in estimating reaction plane.
In this work, we calculate the integrated flow harmonics $v'_n$ and $p_T$-differential flow
harmonics $v_n(p_T)$ with the two-particle cumulant method \cite{Bor01} as,
\begin{equation}
\label{intf}
v'_n\{2\}\!=\!\bigg[\frac{N_n\{2\}}{N_{\rm pair}}\bigg]^{\!\frac{1}{2}}\!\!\!
=\!\bigg\{\!\frac{1}{N_{\rm pair}} \sum_{\alpha=1}^{N_{\rm evt}}\sum_{i\ne j}^{M} \!
\cos n(\phi_i-\phi_j)\bigg\}^{\!\!\frac{1}{2}}\!\!,
\end{equation}
\begin{equation}
\label{diff}
v_n\{2\}(p_T)=\frac{N_n\{2\}(p_T)}{N_{\rm pair}(p_T)} \frac{1}{v'_n\{2\}}\,,
\end{equation}
where $N_{\rm pair}=N_{\rm evt}M(M-1)/2$ is the total number of the particle pairs in
$N_{\rm evt}$ events, $M$ is the particle multiplicity of event, $N_{\rm pair}(p_T)$ and
$N_n\{2\} (p_T)$ are the counts of the particle pairs in the $p_T$ bin with the weights
1 and $\cos[n(\phi_i-\phi_j)]$, respectively.

\begin{figure}[!htbp]
\vspace*{3mm}
\includegraphics[scale=0.70]{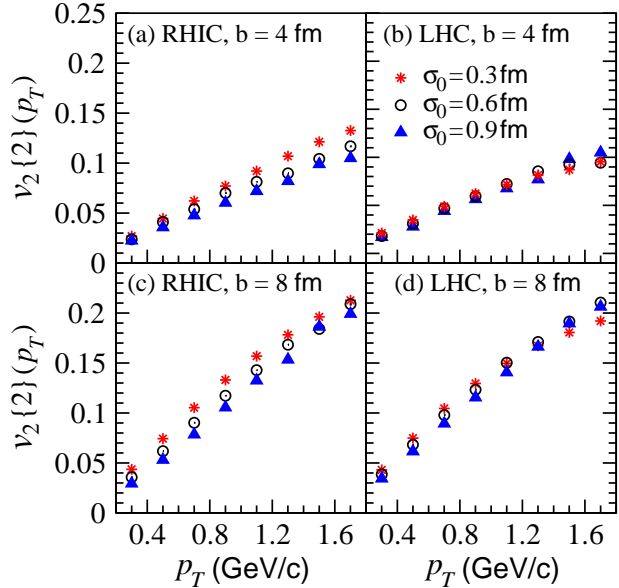}
\caption{(Color online) The pion elliptic flow of the hydrodynamic sources for the Au-Au
collisions at the RHIC energy $\sqrt{s_{NN}}=200$ GeV and the Pb-Pb collisions at the LHC
energy $\sqrt{s_{NN}} =2.76$ TeV with impact parameters $b=$ 4 and 8 fm.  The star, circle,
and triangle symbols are for the parameter $\sigma_0=$ 0.3, 0.6 and 0.9 fm, respectively. }
\label{zfRLv2}
\end{figure}

In Fig. \ref{zfRLv2} we plot the pion $p_T$-differential elliptic flow of the hydrodynamic
sources for the Au-Au collisions at the RHIC energy $\sqrt{s_{NN}}=200$ GeV and the Pb-Pb
collisions at the LHC energy $\sqrt{s_{NN}} =2.76$ TeV with impact parameters $b=$ 4 and 8 fm.
In the calculations, the number of events $N_{\rm evt}$ is six thousand and the particle-pair
number for each event is taken to be 10$^6$.  One can see that the values of elliptic flow 
decrease with increasing $\sigma_0$ for the RHIC sources, and are almost independent of 
$\sigma_0$ for the LHC sources.

\begin{figure}[!htbp]
\vspace*{3mm}
\includegraphics[scale=0.63]{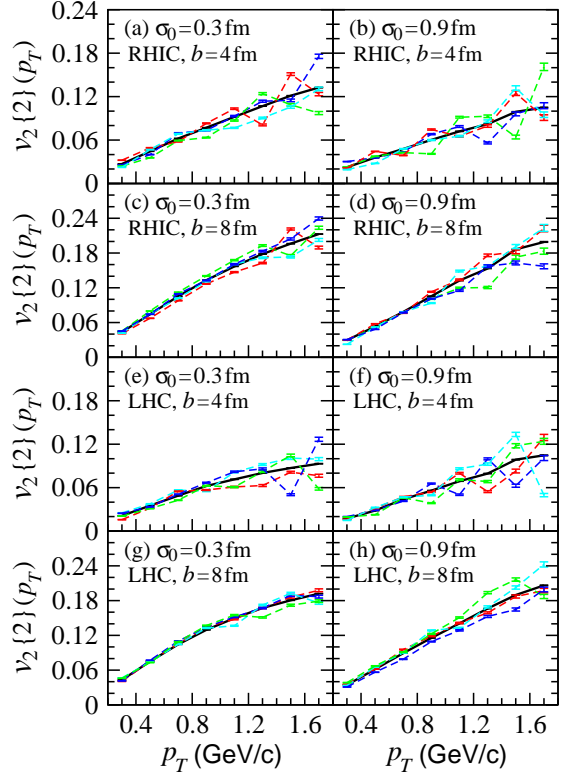}
\vspace*{0mm}
\caption{(Color online) The pion elliptic flow of the event subcollections (dashed lines) each
of them with 100 events, for the RHIC and the LHC sources with $b=$ 4 and 8 fm, $\sigma_0=$ 0.3,
0.6, and 0.9 fm.  The number of particle pair for each event is $10^6$.  The solid lines are
the average results over sixty the subcollections. }
\label{zfRLv2_sb}
\end{figure}

For the hydrodynamic sources with the FIC, the elliptic flow and triangular flow of single
event vary event to event greatly.  These event-by-event fluctuations are associated with 
the FIC as well as the sample statistics.  To reduce the effect of sample statistics on the 
fluctuations, we will examine the elliptic flow and triangular flow of the event subcollections, 
each of them has enough events.  In Fig. \ref{zfRLv2_sb} the dashed lines show the pion elliptic 
flow of the event subcollections each of them with 100 events, for the RHIC and the LHC sources 
with different impact parameter $b$ and $\sigma_0$ values.  The number of particle pair for each 
event is $10^6$.  The solid lines are the results averaged over sixty the subcollections.  The 
error bars of the flow harmonics of the event subcollections are calculated based on the 
definition (\ref{diff}) and the statistical counts, by
\begin{eqnarray}
\label{error}
&&\Delta[v_n\{2\}(p_T)]=\Delta\!\!\left[\frac{N_n\{2\}(p_T)}{N_{\rm pair}(p_T)}\right]
\frac{1}{v'_n\{2\}}+\frac{N_n\{2\}(p_T)}{N_{\rm pair}(p_T)}\, \Delta\!\!\left[\frac{1}{v'_n\{2\}}
\right]\nonumber\\
&&\hspace*{5mm}=v_n\{2\}(p_T) \Bigg\{\frac{\Delta \left[N_n\{2\}(p_T)\right]}{N_n\{2\}(p_T)}
+\frac{\Delta \left[N_{\rm pair}(p_T)\right]}{N_{\rm pair}(p_T)} +\frac{\Delta \left[ N_n\{2\}
\right]}{2N_n\{2\}} +\frac{\Delta N_{\rm pair}}{2N_{\rm pair}}\Bigg\}\nonumber\\
&&\hspace*{5mm}=v_n\{2\}(p_T) \Bigg[\frac{1}{\sqrt{N_n\{2\}(p_T)}} +\frac{1}{\sqrt{N_{\rm pair}
(p_T)}} + \frac{1}{2\sqrt{N_n\{2\}}} + \frac{1}{2\sqrt{N_{\rm pair}}}\Bigg]\nonumber\\
&&\hspace*{5mm} \approx v_n\{2\}(p_T) \Bigg[\frac{1}{\sqrt{N_n\{2\}(p_T)}}
+\frac{1}{\sqrt{N_{\rm pair}(p_T)}} \Bigg]\,.
\end{eqnarray}
One can see from Fig. \ref{zfRLv2_sb} that the elliptic flow calculated with the 100 events
are still with great fluctuations.  They are larger than the statistical errors.  So, the
fluctuations of the flow of event subcollections reflect the intrinsic properties of the
sources.

The azimuthal flow harmonics are related to the initial eccentricities of the source
\cite{AlvRol10,LXHan11},
\begin{equation}
\label{eccen}
\varepsilon_n=\frac{\sqrt{\{\rho^2\cos(n\phi')\}^2 +\{\rho^2\sin(n\phi')\}^2}}{\{\rho^2\}},
\end{equation}
where $\phi'$ is the azimuthal angle related to the reaction plane ($xz$ plane), $\bqr=(x,y)$
is the transverse coordinate of initial source point, and $\{\cdots\}$ denotes the average
over initial source.  In order to give an approximate estimation of the elliptic flow for
the FIC sources, we use a simple distribution of the initial source with separated droplets
to calculate the eccentricity $\varepsilon_2$.  The single-event transverse distribution of
the initial granular source is given by \cite{WonZha04,YanZha09}
\begin{equation}
\label{disdrop}
D(\bqr)=\frac{1}{N_d(2\pi a^2)}\sum_{i=1}^{N_d}\exp\bigg[-\frac{(\bqr-\textbf{\emph{R}}_{\perp
i})^2}{2 a^2}\bigg] ,
\end{equation}
where $N_d$ is the number of droplet, $\textbf{\emph{R}}_{\perp i}=(X_i\,,Y_i)$ are the transverse
coordinates of droplet centers, and $a$ is the standard deviation of the Gaussian distribution.
So, for a single event we have
\begin{eqnarray}
\label{eccen2c1}
\{\rho^2\cos(2\phi')\}=\frac{1}{N_d}\sum_{i=1}^{N_d}(Y_i^2-X_i^2)\,,~~~~~~
\{\rho^2\sin(2\phi')\}=\frac{1}{N_d}\sum_{i=1}^{N_d}2X_iY_i\,,
\end{eqnarray}
\vspace*{-5mm}
\begin{eqnarray}
\label{eccen2c2}
\{\rho^2\}=\frac{1}{N_d}\sum_{i=1}^{N_d}(2a^2+X_i^2+Y_i^2)\,.
\end{eqnarray}
Further, assume that the central coordinates of the droplets obey the Gaussian distribution,
$P(X_i,Y_i)\sim \exp(-X_i^2/2{\cal R}_x^2 -Y_i^2/2{\cal R}_y^2)$.  By substituting the summation
$\frac{1}{N_d}\sum_i$ in Eqs. (\ref{eccen2c1}) and (\ref{eccen2c2}) with $\int dX_idY_i P(X_i,
Y_i)$ and completing the integrations, we get
\begin{eqnarray}
\{\rho^2 \cos(2\phi')\}={\cal R}_y^2 -{\cal R}_x^2\,,~~~~\{\rho^2\sin(2\phi')\}=0\,,~~~~
\{\rho^2\}=2a^2+{\cal R}_x^2+{\cal R}_y^2\,,
\end{eqnarray}
and the $\varepsilon_2$ for a huge number of events is
\begin{equation}
\label{eccen2a}
\varepsilon_{2}=\frac{{\cal R}_y^2 -{\cal R}_x^2}{2a^2+{\cal R}_x^2+{\cal R}_y^2}\,.
\end{equation}
The initial eccentricity decreases with increasing $a$, and the effect of the droplet radius
$a$ on $\varepsilon_2$ becomes slight when ${\cal R}_x$ and ${\cal R}_y$ are large.  These
conclusions are consistent with the results in Fig. \ref{zfRLv2}, where $\sigma_0\sim a$, the
values of elliptic flow decrease with increasing $\sigma_0$ for the RHIC sources, and the
values of elliptic flow for the LHC sources are almost independent of $\sigma_0$ because the
LHC sources have larger initial source sizes.  In Eqs. (\ref{eccen2c1}) and (\ref{eccen2c2}),
$N_d$, $X_i$, and $Y_i$ vary chaotically event to event.  It leads to the fluctuations of the
elliptic flow of single event and the event subcollections with finite number of events as
shown in Fig. \ref{zfRLv2_sb}.

\begin{figure}[!htbp]
\vspace*{3mm}
\includegraphics[scale=0.63]{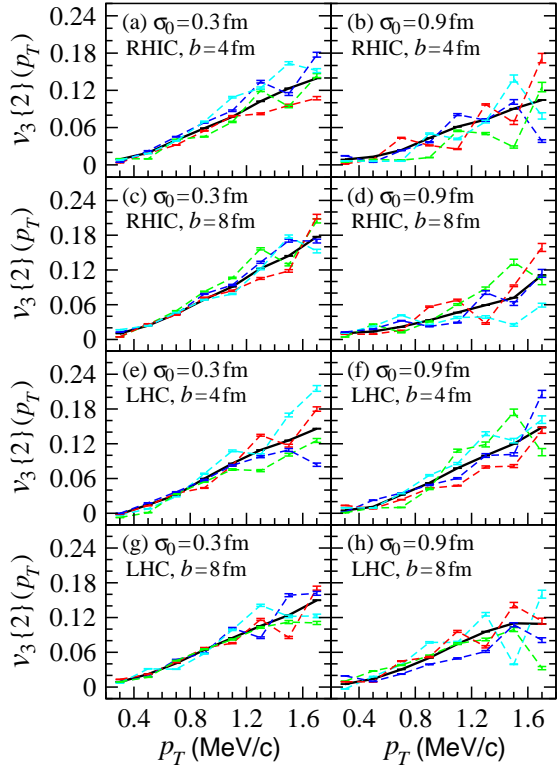}
\vspace*{0mm}
\caption{(Color online) The pion triangular flow of the event subcollections (dashed lines) as
in Fig. \ref{zfRLv2_sb}.  The solid lines are the average results over sixty the subcollections. }
\label{zfRLv3_sb}
\end{figure}

In Fig. \ref{zfRLv3_sb} the dashed lines show the triangular flow for the event subcollections
as in Fig. \ref{zfRLv2_sb}.  The solid lines are the results averaged over sixty the
subcollections.  As compared to the results of elliptic flow in Fig. \ref{zfRLv2_sb}, the
fluctuations of triangular flow are larger.  The values of triangular flow are related to the
initial eccentricity $\varepsilon_3$, which becomes nonzero due to the granular inhomogeneous
structure of the initial sources.  Because involving elliptic integrations, we cannot obtain 
an analytic expression of $\varepsilon_3$ even for the simple droplet model.

\begin{table*}
\begin{center}
\caption{The values of $\sigma_0$ (column 2), eccentricities (columns 3 and 6), average flow 
harmonics columns 4 and 7), scaled average flow harmonics (columns 5 and 8), average transverse 
diameter (columns 9), and granularity length (column 10) of the initial sources for the RHIC 
and LHC collisions with impact parameters $b=$ 4 and 8 fm. }
\begin{tabular}{l|cccccccccc}
\hline\hline
~~&~~$\sigma_0\text{(fm)}$~~&~~$\varepsilon_2$&~\,$\langle v_2\{2\}(p_T)\rangle$\,&~
${\widetilde v}_2$~&~~~~\,$\varepsilon_3$~\,&~$\langle v_3\{2\}(p_T)\rangle$&~~${\widetilde
v}_3$~~&~~$2\{\rho\}$(fm)&~$L_{\xi}$(fm)~\\
\hline
~~RHIC~ ~ & 0.3 & ~0.189 & 0.082 & 0.434 & \,~~~~0.091 & 0.071 & 0.780 & ~~6.57 & 6.80\\
~$b=4$ fm~& 0.6 & ~0.183 & 0.072 & 0.393 & \,~~~~0.085 & 0.061 & 0.718 & ~~6.68 & 7.67\\
~         & 0.9 & ~0.172 & 0.065 & 0.378 & \,~~~~0.076 & 0.052 & 0.684 & ~~6.87 & 9.47\\
\hline
~~RHIC~~  & 0.3 & ~0.375 & 0.137 & 0.365 & \,~~~~0.143 & 0.085 & 0.594 & ~~5.20 & 5.49\\
~$b=8$ fm~& 0.6 & ~0.356 & 0.126 & 0.351 & \,~~~~0.129 & 0.062 & 0.480 & ~~5.37 & 6.71\\
~         & 0.9 & ~0.322 & 0.117 & 0.363 & \,~~~~0.110 & 0.045 & 0.409 & ~~5.62 & 9.53\\
\hline
~~LHC~~   & 0.3 & ~0.172 & 0.063 & 0.366 & \,~~~~0.076 & 0.071 & 0.934 & ~~7.30 & 7.50\\
~$b=4$ fm~& 0.6 & ~0.167 & 0.062 & 0.371 & \,~~~~0.075 & 0.071 & 0.947 & ~~7.40 & 8.27\\
~         & 0.9 & ~0.165 & 0.062 & 0.376 & \,~~~~0.065 & 0.068 & 1.046 & ~~7.58 & 9.79\\
\hline
~~LHC~~   & 0.3 & ~0.367 & 0.131 & 0.357 & \,~~~~0.112 & 0.075 & 0.670 & ~~5.81 & 6.08\\
~$b=8$ fm~& 0.6 & ~0.350 & 0.130 & 0.371 & \,~~~~0.105 & 0.071 & 0.676 & ~~5.97 & 7.11\\
~         & 0.9 & ~0.325 & 0.125 & 0.385 & \,~~~~0.092 & 0.061 & 0.663 & ~~6.19 & 9.35\\
\hline\hline
\end{tabular}
\end{center}
\end{table*}

In the third and sixth columns of Table I, we list the values of $\varepsilon_2$ and
$\varepsilon_3$ calculated in Eq. (\ref{eccen}) by averaging with the initial distributions
of energy density (see the top panels in Fig. \ref{zf-hydro}) of the six thousand events for
the RHIC and LHC sources with the different $b$ and $\sigma_0$ values, respectively.  The
values of $\varepsilon_2$ and $\varepsilon_3$ increase with increasing $b$ and decrease
with increasing $\sigma_0$ for fixed $b$.  The columns 4 and 7 of Table I list the average
values of $v_2\{2\}(p_T)$ and $v_3\{2\}(p_T)$ over $p_T$ and sixty the subcollections
each of them with 100 events.  For fixed $b$, the variations of $\langle v_2\{2\}(p_T)
\rangle$ and $\langle v_3\{2\}(p_T)\rangle$ with $\sigma_0$ are consistent with the
variations of $\varepsilon_2$ and $\varepsilon_3$ with $\sigma_0$, respectively.
However, the values of $\langle v_3\{2\}(p_T)\rangle$ and $\varepsilon_3$ for the same
$\sigma_0$ but different $b$ are not completely consistent.  For instance, the value of
$\varepsilon_3$ for the RHIC source with $\sigma_0=0.9$ fm and $b=4$ fm is 0.076 and
smaller than the value 0.110 for the RHIC source with the same $\sigma_0$ and $b=8$ fm,
but the corresponding $\langle v_3\{2\}(p_T)\rangle$ value for the former source is 0.052
and larger than the $\langle v_3\{2\}(p_T)\rangle$ value 0.045 for the later source.
This contradiction of the values of $\langle v_3\{2\}(p_T)\rangle$ and $\varepsilon_3$
for different $b$ indicates that there exist other effects on the triangular flow values
for different impact parameter.  In the fifth and eighth columns of Table I, we present
the ratios ${\widetilde v}_2 \equiv\langle v_2\{2\}(p_T)\rangle/\varepsilon_2$ and
${\widetilde v}_3\equiv\langle v_3\{2\}(p_T) \rangle/\varepsilon_3$, respectively.
One can see that the values of the scaled average elliptic flow and triangular flow, 
${\widetilde v}_2$ and ${\widetilde v}_3$, are insensitive to $\sigma_0$.

\subsection{Fluctuations of the pion flow harmonics and HBT correlation functions of event
subcollections}
From Figs. \ref{zfRLv2_sb} and \ref{zfRLv3_sb} it can be seen that the elliptic flow and
triangular flow of the event subcollections (dashed lines) are fluctuated.
These fluctuations are smoothed out in the observables for all the event subcollections
(solid lines).  To display the flow fluctuations of the event subcollections in the analysis
for all events (or all event subcollections), we introduce the distribution $dN/df$ of the
differences,
\begin{equation}
f_{vn}=\big|v_n^{(i)}\{2\}(p_T)-v_n^{(j)}\{2\}(p_T)\big|, ~~~~(j\ne i),
\end{equation}
accumulating for all the event subcollections and the $p_T$ bins in the considered region.
Here, the superscript denotes the event subcollection.

\begin{figure}[!htb]
\vspace*{0mm}
\includegraphics[scale=0.63]{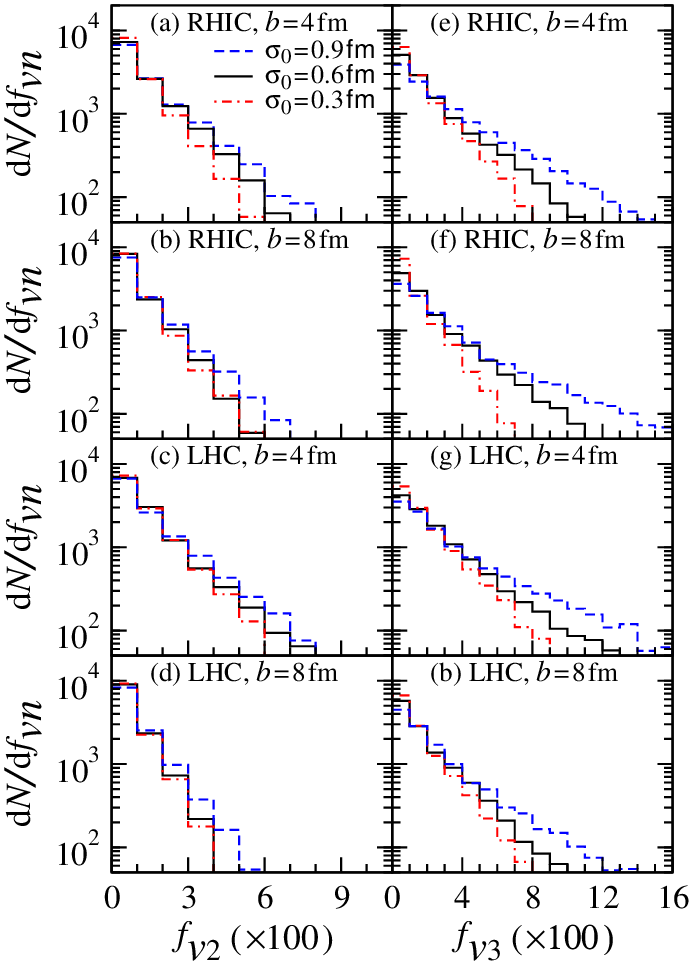}
\vspace*{0mm}
\caption{(Color online) The distributions $dN/df_{vn}$ for the RHIC and the LHC sources with
different impact parameter $b$ and $\sigma_0$. }
\label{zfRLdnfb48}
\end{figure}

Unlike some cumulate quantities which smooth out the fluctuations in the analysis for all
events, $dN/df$ is a fluctuation distribution.  It becomes wide for the variables $v_n^{(i)}
\{2\}$ with large fluctuations, and the similar analysis was used in the investigations of
the fluctuations of single-event HBT correlation functions \cite{Zha05,RenZha08,HuZha13}.  
Because the event number of collision is very huge (in principle it may reach any large 
number if prolonging experiment time), the number of the event subcollections will be large 
enough to overcome the influence of statistic fluctuations on the distributions.  In Fig.
\ref{zfRLdnfb48}, we plot the distributions $dN/df_{vn}$ for the RHIC and LHC sources
with different impact parameter $b$ and $\sigma_0$ values.  The widths of the distributions
increase with increasing $\sigma_0$, and in most case the widths decrease with increasing
$b$ for fixed $\sigma_0$.  The distributions of $dN/df_{v3}$ are wider as compared to the
corresponding $dN/df_{v2}$ distributions, and the width of the distribution of triangular
flow is sensitive to $\sigma_0$.  

The wider the distributions $dN/df_{vn}$, the larger the fluctuations are.  The width of the
distribution reflects the fluctuation magnitude of the flow harmonics.  The changes of the 
widths of the distributions $dN/df_{v2}$ and $dN/df_{v3}$ with impact parameter $b$ and 
$\sigma_0$ reflect the variations of the fluctuation magnitudes with the granular inhomogeneity 
of the initial sources.  In order to quantify the granular inhomogeneity of the initial source, 
we introduce a granularity length of the initial source as the product of the initial transverse 
radius ${\cal R}_{\perp}$ and the transverse granularity parameter $\xi_{\perp}$ \cite{YanZha09} 
of the initial source,
\begin{eqnarray}
\label{Lxi}
L_{\xi}={\cal R}_{\perp}\xi_{\perp}={\cal R}_{\perp}\frac{({\cal R}_{\perp}/\sigma_0)^2}{N_d-2}
\approx\frac{2\{\rho\}^3}{\{\rho\}^2-4\sigma_0^2}, ~~~~~~(2\sigma_0<\{\rho\}).
\end{eqnarray}
Here, we replace ${\cal R}_{\perp}$ approximately with the average transverse radius of the
initial source $\{\rho\}$, and replace $N_d$ approximately with $\frac{1}{2}(\{\rho\}/
\sigma_0)^2$, considering approximately the same numbers of the hot spot and cold valley in
the initial source.  Clearly, the concept of granular source requires $N_d\geq2$.  So we have
$2\{\rho\} <L_{\xi}<\infty$.  From Eq. (\ref{Lxi}), the granularity length increase when the
initial source radius $\{\rho\}$ (associated with collision impact parameter and energy)
increases.  For fixed $\{\rho\}$, $L_{\xi}$ increases with increasing $\sigma_0$ because the
droplet number decreases with increasing $\sigma_0$.

In the right two columns of Table I, we presents the values of $2\{\rho\}$ (calculated with
the initial energy density of source) and $L_{\xi}$ for the RHIC and the LHC sources respectively.
The values of $\{\rho\}$ decrease with increasing impact parameter $b$ and increase slightly with
increasing $\sigma_0$ for fixed $b$.  The values of $\{\rho\}$ for the LHC sources are larger
than the corresponding results for the RHIC sources.  In most case, the value of $L_{\xi}$ for 
smaller $b$ is larger than that for larger $b$ with the same $\sigma_0$, because the initial 
source radius is larger for smaller $b$.  However, for $\sigma_0=0.9$ fm, the $L_{\xi}$ value 
for the RHIC source with $b=8$ fm is larger than that for $b=4$ fm.  This is because that the
effect of droplet number on the granularity length becomes important when $(\{\rho\}/
\sigma_0)^2$ is small.

From Fig. \ref{zfRLdnfb48} and the $L_{\xi}$ values in Table I, we observe that the widths of
the distributions $dN/df_{v2}$ and $dN/df_{v3}$ become wider as $L_{\xi}$ increases, for the
RHIC sources or the LHC sources.  This is obvious for the triangular flow distributions.  We
conclude that the fluctuation magnitude of the triangular flow of event subcollections is
sensitively dependent on the granularity length of the initial source.  For the sources of 
the same collision energy, the fluctuation magnitude of the triangular flow increases with 
$L_{\xi}$ monotonically.  However, for different collision energies, we observe the variation 
of the distribution width with $L_{\xi}$ is not completely in a monotonic manner.  For instance, 
the $L_{\xi}$ values for the LHC source with $b=8$ fm and $\sigma_0=0.6$ fm is 7.11 fm and 
larger than the $L_{\xi}$ value 6.71 fm for the RHIC source with the same $b$ and $\sigma_0$, 
but the distribution width for the LHC source is smaller than that for the RHIC source.
This indicates the limitation of $L_{\xi}$ for the collisions with greatly different energies.

\begin{figure}[!htb]
\vspace*{0mm}
\includegraphics[scale=0.63]{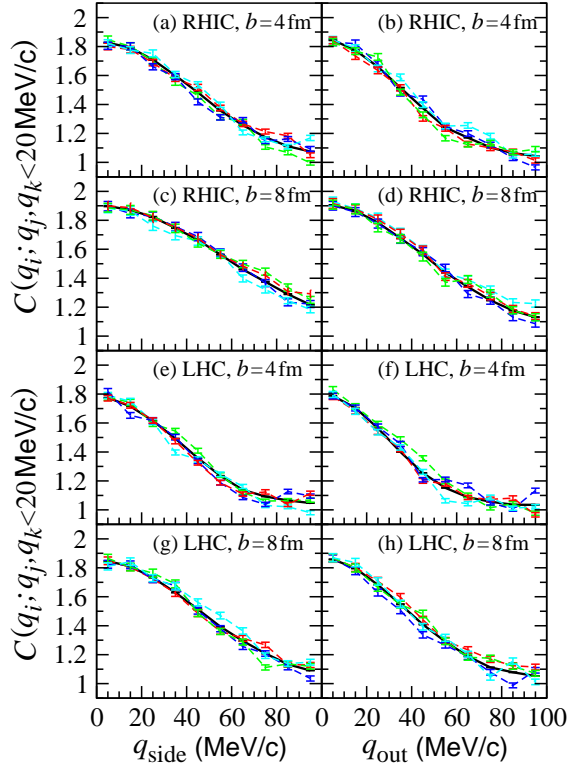}
\vspace*{0mm}
\caption{(Color online) The HBT correlation functions of the event subcollections (dashed lines)
each of them with the 10 events for the RHIC and the LHC sources with different impact parameters 
and $\sigma_0=0.6$ fm.  For each event the particle pair number is $10^6$.  The solid lines are 
the average results over six hundred the subcollections. }
\label{zfRLCqso}
\end{figure}

\begin{figure}[!htb]
\includegraphics[scale=0.60]{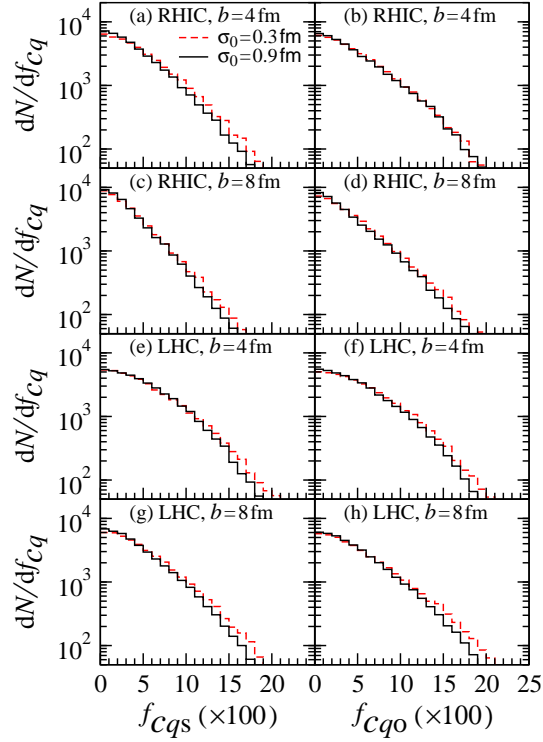}
\caption{(Color online) The distributions $dN/df_{Cq}$ for the RHIC and the LHC sources
with different values of impact parameter $b$ and $\sigma_0$. }
\label{zfRLCdnf}
\end{figure}

For granular sources, the single- or several-event HBT correlation functions are fluctuated
\cite{WonZha04,Zha05,RenZha08,HuZha13}.  We plot in Fig. \ref{zfRLCqso} the two-pion HBT
correlation functions of the event subcollections (dashed lines) each of them with 10 the
events for the RHIC and the LHC sources with different impact parameters and $\sigma_0=
0.6$ fm.  Here, $q_{\rm side}$ and $q_{\rm out}$ are the relative transverse momenta of 
the pion pair in the ``side" and ``out" directions \cite{Ber88,Pra90}.  The solid lines 
are the results averaged over six-hundred the subcollections.  We observe the fluctuations 
of the HBT correlations of the event subcollections, and the fluctuations are smoothed out 
in the average results.  As the analyses for the flow fluctuations, we investigate the 
distribution, $dN/df_{Cq}$, for the HBT correlation functions $C(q_{\rm side})$ and 
$C(q_{\rm out})$.  In Fig. \ref{zfRLCdnf}, we plot the distributions of $dN/df_{Cq}$ 
the RHIC and the LHC sources with different $b$ and $\sigma_0$ values.  We observe that 
the distributions are insensitive to $\sigma_0$, although they become wider as compared 
to the distributions for the source with smoothed initial conditions \cite{RenZha08,HuZha13}.  
Unlike elliptic flow and triangular flow, HBT correlation functions reflect more about the 
source freeze-out geometry and dynamics rather than the initial details of the sources.

\begin{figure}[!htb]
\includegraphics[scale=0.60]{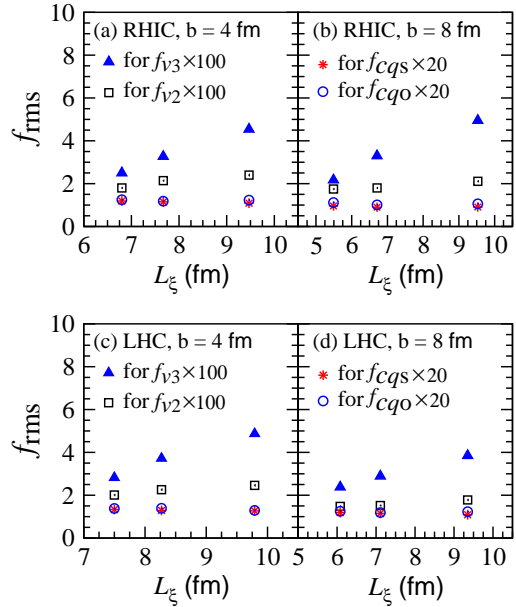}
\caption{(Color online) The variations of the root-mean-square $f_{\rm rms}$ of the elliptic
flow, triangular flow, and HBT correlation functions with $L_{\xi}$, for the RHIC and LHC
sources with $b=$ 4 and 8 fm. }
\label{zfRLrms}
\end{figure}

The width of the distribution $dN/df$, thus the fluctuation magnitude, can be described
by the root-mean-square (RMS), $f_{\rm rms}$, quantitatively.  Finally, we plot in Fig.
\ref{zfRLrms} the variations of the RMS of the elliptic flow, triangular flow, and HBT
correlation functions with the granularity length of the initial source.  One can see
that the RMS values of the flow harmonics increase as $L_{\xi}$ increases.  But the RMS
values of the HBT correlation functions are almost independent of $L_{\xi}$.  The RMS
values of the triangular flow are very sensitive to $L_{\xi}$.
This dependence provide a way to study the granular inhomogeneity of the initial sources 
through the analyses of the fluctuations of triangular flow in ultrarelativistic heavy 
ion collisions.

\section{Summary and discussions}
We investigate the fluctuations of pion elliptic flow, triangular flow,
and HBT correlation functions for the hydrodynamic sources of the Au-Au collisions 
at $\sqrt{s_{NN}}=200$ GeV and the Pb-Pb collisions at $\sqrt{s_{NN}}=2.76$ TeV.  
The initial sources are generated by the HIJING, and then evolve described by ideal 
relativistic hydrodynamics in (2+1) dimensions with Bjorken's longitudinal boost 
invariance.  The EOS of s95p-PCE, which combines the hadron resonance gas at low 
temperature and the lattice QCD results at high temperature, is employed in the 
hydrodynamic calculations.  For the hydrodynamic sources with the FIC, the
elliptic flow, triangular flow, and HBT correlation functions are fluctuated 
event-by-event.  These FIC-caused fluctuations survive in the observables obtained 
with event subcollections.  To display the fluctuations we introduce the fluctuation 
distribution, $dN/df$, of the observable of event subcollections, which becomes wide 
for the observable with large fluctuations.  We also introduce the granularity length 
$L_{\xi}$ to describe the granular inhomogeneity of the initial sources.  
The relationships between the granularity length and the fluctuations of the pion 
elliptic flow, triangular flow, and HBT correlation functions of event subcollections 
are investigated.  Our investigations indicate that the FIC lead to event-by-event 
fluctuations of the elliptic flow, triangular flow, and HBT correlation functions.  
These FIC-caused fluctuations can be detected by the fluctuation distributions of the 
observables of event subcollections.  The fluctuations of the triangular flow of event 
subcollections are sensitive to the granularity length of the initial source.  
This dependence provide a way to investigate the granular inhomogeneity of the initial
sources through analysing the fluctuations of triangular flow in ultrarelativistic
heavy ion collisions.  

For the initial source with fluctuating matter distribution, the initial velocities 
of fluid-cells in the source is usually nonzero and also fluctuated.  Considering the 
very low ratio of shear viscosity to entropy density of the QGP matter and the 
stability of the hydrodynamic evolution of the FIC sources with nonzero initial 
velocities of fluid-cells in the source, we use an ideal hydrodynamics model in 
this work.  On the other hand, we also did not consider the initial fluctuation 
in longitudinal direction in this work.  Further investigations on the relationship 
between the source initial granular inhomogeneity and the fluctuations of final 
observables for different initial source models and based on viscous hydrodynamic 
description of source evolution will be of great interest.

\begin{acknowledgments}
We thank Dr. Longgang Pang and Dr. Luan Cheng for helpful discussions.  This work is supported
by the National Natural Science Foundation of China under Contract No. 11275037.
\end{acknowledgments}

\end{document}